\RequirePackage{graphicx}
\documentclass[aps,prl,reprint,showpacs,groupedaddress,amssymb]{revtex4-1}
\usepackage{atlasphysics}
\usepackage[hyperindex,breaklinks]{hyperref} 
\usepackage[hang,center]{subfigure}
\usepackage{dcolumn}
\usepackage{preprintcover}
\PreprintCoverPaperTitle{Search for a supersymmetric partner to the top quark\\ in final
  states with jets and missing transverse momentum\\ at $\bf\sqrt{s}=7$~TeV
  with the ATLAS detector}
\PreprintIdNumber{CERN-PH-EP-2012-201}
\PreprintCoverAbstract{\noindent\fontseries{m}  
A search for direct pair production of supersymmetric top squarks
  (\stopone) is presented, assuming the \stopone\ decays into a top
  quark and the lightest supersymmetric particle, $\ninoone$, and that
  both top quarks decay to purely hadronic final states.  A total of
  $16\;(4)$ events are observed compared to a predicted Standard Model
  background of ${13.5}^{+\,\,3.7}_{-\,\,3.6}\;
  ({4.4}^{+\,\,1.7}_{-\,\,1.3})$ events in two signal regions based on
  \ensuremath{\int\!\!{\cal L}\,{\rm d}t = 4.7\,\ifb} of $pp$
  collision data taken at $\rts=7\tev$ with the ATLAS detector at the
  LHC.  An exclusion region in the \stopone\ versus $\ninoone$ mass
  plane is evaluated: $370<m_{\stopone}<465$~GeV is excluded for
  $m_{\ninoone}\sim0$~\GeV\ while $m_{\stopone}=445$~\GeV\ is excluded
  for $m_{\ninoone}\le 50$~\GeV.}
\PreprintJournalName{Physical Review Letters}
\begin{document}
\title{Search for a supersymmetric partner to the top quark in final
  states with\\ jets and missing transverse momentum at $\bf\sqrt{s}=7$~TeV
  with the ATLAS detector}
\author{The ATLAS Collaboration}
\date{August 7, 2012}
\begin{abstract}\noindent
  A search for direct pair production of supersymmetric top squarks
  (\stopone) is presented, assuming the \stopone\ decays into a top
  quark and the lightest supersymmetric particle, $\ninoone$, and that
  both top quarks decay to purely hadronic final states.  A total of
  $16\;(4)$ events are observed compared to a predicted Standard Model
  background of ${13.5}^{+\,\,3.7}_{-\,\,3.6}\;
  ({4.4}^{+\,\,1.7}_{-\,\,1.3})$ events in two signal regions based on
  \ensuremath{\int\!\!{\cal L}\,{\rm d}t = 4.7\,\ifb} of $pp$
  collision data taken at $\rts=7\tev$ with the ATLAS detector at the
  LHC.  An exclusion region in the \stopone\ versus $\ninoone$ mass
  plane is evaluated: $370<m_{\stopone}<465$~GeV is excluded for
  $m_{\ninoone}\sim0$~\GeV\ while $m_{\stopone}=445$~\GeV\ is excluded
  for $m_{\ninoone}\le 50$~\GeV.
\end{abstract}
\maketitle
The Standard Model (SM) is a successful but incomplete model of
particle interactions.  Supersymmetry
(SUSY)~\cite{Miyazawa:1966,Ramond:1971gb,Golfand:1971iw,Neveu:1971rx,Neveu:1971iv,Gervais:1971ji,Volkov:1973ix,Wess:1973kz,Wess:1974tw}
provides an elegant solution to cancel the quadratic mass divergences
that would accompany a SM Higgs boson by introducing supersymmetric
partners of all SM particles, such as a scalar partner of the top
quark (\stop).  Like \ttbar, direct $\stop\stopstar$ is produced
primarily through gluon fusion at the Large Hadron Collider (LHC).
The production cross section depends mostly on the mass of the top
partner and has minimal dependence on other SUSY
parameters~\cite{Beenakker:1997ut,Beenakker:2010nq,Beenakker:2011fu}.
The LHC enables searches for direct $\stop\stopstar$ production at
higher mass scales than previous
accelerators~\cite{Achard:2003ge,Abdallah:2003xe,Abbiendi:2002mp,Heister:2002hp,Abazov:2012cz,Abazov:2010xm,Aaltonen:2009sf,Abazov:2009ps,Abazov:2008kz,Brigliadori:2008vf,Abazov:2008rc,Aaltonen:2007sw,Acosta:2003ze,Aad:2012cz,ATLAS-CONF-2012-059}.
The viability of SUSY as a scenario to stabilize the Higgs potential
and to be consistent with electroweak
naturalness~\cite{Barbieri:1987fn,Carlos:1993yy} is tested by the
search for \stop\ below the \TeV\ scale.

In this Letter, we present a search for direct $\stop\stopstar$
production assuming $\stopone\rightarrow t\ninoone \rightarrow
bW\ninoone$ where \stopone\ is the lightest \stop\ eigenstate and
\ninoone\ represents the lightest supersymmetric particle (LSP) in
$R$-parity conserving
models~\cite{Fayet:1976et,Fayet:1977yc,Farrar:1978xj,Fayet:1979sa,Dimopoulos:1981zb}.
We consider events where both $W$ bosons decay hadronically, yielding
a final state with six high transverse momentum (\pt) jets from the
all-hadronic \ttbar\ final state and large missing transverse momentum
(\MET) from the LSPs.  The kinematics of both top quarks are therefore
fully specified by the visible decay products.  Additionally, SM
backgrounds from all-hadronic \ttbar\ are suppressed as there is no
intrinsic \MET\ except from semi-leptonic $c$- and $b$-quark decays.
The dominant background consists of \ttbar\ events that contain a
$W\rightarrow\ell\nu$ decay where the lepton ($\ell$), often of $\tau$
flavor, is either lost or mis-identified as a jet and which have large
\MET\ from the neutrino.

The data sample was acquired during 2011 in LHC $pp$ collisions at a
center-of-mass energy of 7~TeV with the ATLAS
detector~\cite{Aad:2008zzm}, which consists of tracking detectors
surrounded by a 2~T superconducting solenoid, calorimeters, and a muon
spectrometer in a toroidal magnetic field.  The high-granularity
calorimeter system, with acceptance covering $|\eta| <
4.9$~\footnote{ATLAS uses a right-handed coordinate system with its
  origin at the nominal interaction point in the center of the
  detector and the $z$-axis along the beam pipe.  Cylindrical
  coordinates $(r,\phi)$ are used in the transverse plane, $\phi$
  being the azimuthal angle around the beam pipe. The pseudorapidity
  is defined in terms of the polar angle $\theta$ as
  $\eta=-\ln\tan(\theta/2)$.}, is composed of liquid argon with lead,
copper, or tungsten absorbers and scintillator tiles with steel
absorbers.  This data set, composed of events with a high-\pt\ jet and
large \MET\ as selected by the trigger system, corresponds to an
integrated luminosity of $4.7$~\ifb\ with a relative uncertainty of
3.9\%~\cite{Aad:2011dr,ATLAS-CONF-2011-116}.

Jets are constructed from three-dimensional clusters of calorimeter
cells using the anti-$k_t$ algorithm with a distance parameter
of~0.4~\cite{Cacciari:2008gp,Cacciari:2005hq}.  Jet energies are
corrected~\cite{Aad:2011he} for losses in material in front of the
active calorimeter layers, detector inhomogeneities, the
non-compensating nature of the calorimeter, and the impact of multiple
overlapping $pp$ interactions. These corrections are derived from test
beam, cosmic ray, and $pp$ collision data, and from a detailed
Geant4~\cite{Agostinelli:2002hh} detector simulation~\cite{:2010wqa}.
Jets containing a $b$-hadron are identified with an algorithm
exploiting both the impact parameter and secondary vertex
information~\cite{ATLAS-CONF-2012-043,ATLAS-CONF-2012-040}.  A factor
correcting for the slight differences in the $b$-tagging efficiency
between data and the Geant4 simulation is applied to each jet in the
simulation.  The $b$-jets are restricted to the fiducial region of the
tracker, $|\eta|<2.5$.  Non-\ttbar\ backgrounds are minimized by
requiring either $\geq1$~$b$-jets with a selection corresponding to a
60\% efficiency with a low $<0.2\%$ mis-identification rate (tight),
or $\geq2$~$b$-jets each with 75\% efficiency but a higher
$\approx1.7\%$ mis-identification rate per $b$-jet (loose).

The \MET\ is the magnitude of \ptmiss, the negative vector sum of the
\pt\ of the clusters of calorimeter cells, calibrated according to
their associated reconstructed object (e.g., jets and electrons), and
of the \pt\ of muons above $10$~\GeV\ within $|\eta|<2.4$.  Events
containing fake \MET\ induced by jets associated with calorimeter
noise or non-collision backgrounds~\cite{ATLAS-CONF-2012-020}, or by
cosmic-ray muons or poorly reconstructed
muons~\cite{ATLAS-CONF-2011-063,Aad:2010mr}, are removed from
consideration.  Large \ptmiss\ colinear with a high-\pt\ jet could
indicate a significant fluctuation in the reconstructed jet energy or
the presence of a semi-leptonic $c$- or $b$-quark decay.  Therefore,
the difference in azimuthal angle ($\Delta\phi$) between the \ptmiss\
and any of the three highest-\pt\ jets in the event,
$\Delta\phi(\ptmiss,{\rm jet})$, is required to be $>\pi/5$ radians.
Fake \MET\ is also suppressed by requiring that the $\Delta\phi$
between the above computed \ptmiss\ and one calculated with the
tracking system, using tracks having $\pt>0.5$~GeV, is $<\pi/3$
radians.

Events are required to have at least one jet with $\pt>130$~\GeV\ in
$|\eta|<2.8$ and $\MET>150$~\GeV\ to ensure full efficiency of the
trigger.  At least five other jets having $\pt>30$~\GeV\ and
$|\eta|<2.8$ must be present.  In addition to the jet and \MET\
requirements, events containing ``loose''
electrons~\cite{Aad:2011mk,ATLAS-COM-PHYS-2011-1669} with
$\pt>20$~\GeV\ and $|\eta|<2.47$ that do not overlap with any jet
within $\Delta R<0.4$, where $\Delta
R=\sqrt{(\Delta\eta)^2+(\Delta\phi)^2}$, are rejected.  Similarly,
events with muons~\cite{ATLAS-CONF-2011-063,ATLAS-CONF-2011-021}
having $\pt>10$~\GeV\ and $|\eta|<2.4$ that are separated by $\Delta
R>0.4$ from the nearest jet are rejected.  A jet with 1--4 tracks and
$\Delta\phi(\ptmiss,{\rm jet}) < \pi/5$ indicates a likely
$W\rightarrow\tau\nu$ decay. Events with such $\tau$-like jets that
have transverse mass $\mt=\sqrt{2\pt\MET(1-\cos\Delta\phi)}<100\gev$
are rejected.

The presence of high-\pt\ top quarks that decay through $t\rightarrow
bW\rightarrow bjj$ in the \stoppair\ final state is exploited to
further reduce SM backgrounds by only considering events with
reconstructed three-jet invariant masses consistent with the top-quark
mass ($m_{t}$).  A clustering technique resolves the combinatorics
associated with high-multiplicity jet events.  The three closest jets
in the $\eta-\phi$ plane are combined to form one triplet; a second
triplet is formed from the remaining jets by repeating the procedure.
The resulting three-jet mass ($m_{jjj}$) spectrum is shown in
Fig.~\ref{fig:topmass} for the control region constructed from
$\ell+$jets events (defined below).  There is a clear peak associated
with the hadronically-decaying top quarks above a small non-\ttbar\
background.  A requirement of $80<m_{jjj}<270$~\GeV\ is placed on each
reconstructed triplet in the event.  The kinematics of the
$t\rightarrow bW\rightarrow b\ell\nu$ decay is also exploited to
further reduce the dominant $\ell+$jets \ttbar\ background, as the
\mt\ distribution of the \ptmiss\ and $b$-jet (\mtj) has an endpoint
at $m_{t}$.  When there are $\geq2$~loose $b$-jets, the \mtj\ for the
$b$-jet closest to the \ptmiss\ is required to be $>175$~\GeV.  The
largest \mtj, calculated for each of the four highest-\pt\ jets, is
required to be $>175$~\GeV\ in the case of only one tight $b$-jet.

\begin{figure}[t]
\includegraphics[width=\columnwidth,trim=14 0 8 20,clip=true]{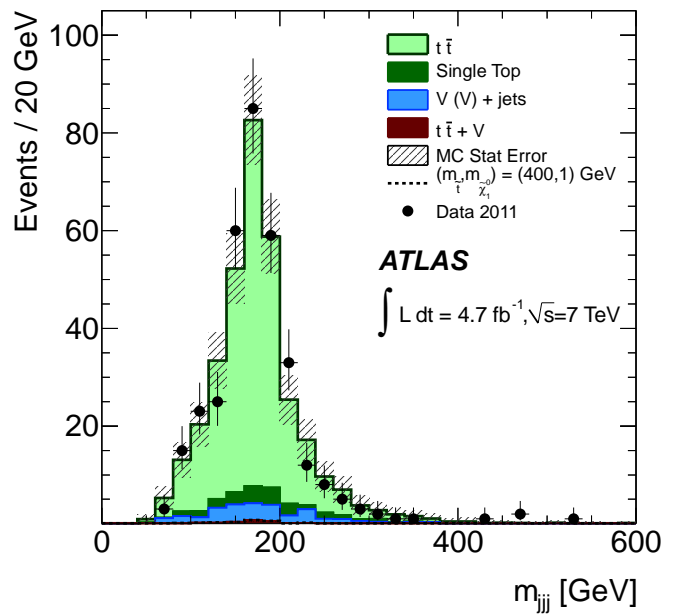}
\caption{\label{fig:topmass} Three-jet invariant mass distribution of
  the hadronic top-quark candidate in the control region constructed
  from $\ell+$jets events.  Data are indicated by points; shaded
  histograms represent contributions from several SM sources (with
  \ttbar\ scaled by $0.66$).  The hatched error bars indicate the
  total statistical uncertainty on the expected background.  The
  distribution for the $m_{\stopone} = 400\gev, m_{\ninoone} = 1\gev$
  signal expectation in this control region is overlaid.  }
\end{figure}
Two signal regions (SR) are defined including the above kinematic and
mass requirements.  The first, which requires $\MET>150$~\GeV\ (SRA),
is optimized for low $m_{\stopone}$, while the second, requiring
$\MET>260$~\GeV\ (SRB), is used for higher $m_{\stopone}$.  Using
these signal regions, the search is most sensitive to \stoppair\
production with $350\lesssim m_{\stopone}\lesssim 500$~GeV and
$m_{\ninoone} \muchless m_{\stopone}$.  Signal events are simulated
using Herwig++~\cite{Bahr:2008pv} with the
MRST2007LO*~\cite{Sherstnev:2007nd} parton distribution functions
(PDF) generated with the \stopone\ and \ninoone\ masses at fixed
values in a grid with 50~GeV spacing. The mixing parameters are chosen
such that the lightest scalar top is mostly the partner of the
right-handed top quark. The branching fraction of $\stopone\rightarrow
t\ninoone$ is set to 100\%~\footnote{All simulations of top-quark
  related processes assume a top-quark mass of 172.5~\GeV.}.  Signal
cross sections are calculated to next-to-leading order in the strong
coupling constant, including the resummation of soft gluon emission at
next-to-leading-logarithmic accuracy
(NLO+NLL)~\cite{Beenakker:1997ut,Beenakker:2010nq,Beenakker:2011fu}. The
nominal production cross section and associated uncertainty are taken
from an envelope of cross section predictions using different PDF sets
and factorization and renormalization scales, as described in
Ref.~\cite{Kramer:2012bx}.  The \stoppair\ cross section for
$m_{\stopone}=400$~GeV is $\sigma_{\stoppair}=0.21\pm0.03$~pb.

In the signal region, the dominant source of SM background is
$\ttbar\rightarrow\tau+$jets events where the $\tau$ lepton is
reconstructed as a jet.  Additional, smaller, backgrounds include
other $\ttbar\to\ell+$jets final states, $\ttbar+V$ where $V$
represents a $W$ or $Z$ boson, single top quark production, $V+$jets,
and $VV$+jets. The \ttbar\ events are produced with
ALPGEN~\cite{Mangano:2002ea} using the CTEQ6L1
PDF~\cite{Pumplin:2002vw} and interfaced to
HERWIG~\cite{Corcella:2000bw,ATLAS-PHYS-PUB-2011-008} for particle
production and JIMMY~\cite{Butterworth:1996zw} for the underlying
event model.  Additional \ttbar\ samples generated with
MC@NLO~\cite{Frixione:2002ik,Frixione:2003ei} and
AcerMC~\cite{Kersevan:2004yg}, interfaced to HERWIG and JIMMY, are
used to estimate event generator systematic uncertainties.  Samples of
$\ttbar+V$ are produced with MadGraph~\cite{Alwall:2007st} interfaced
to
PYTHIA~\cite{Sjostrand:2006za,ATLAS-PHYS-PUB-2011-008,ATLAS-PHYS-PUB-2011-014}.
Single top events are generated with
MC@NLO~\cite{Frixione:2005vw,Frixione:2008yi} and AcerMC.  The
associated production of $W$ and $Z$ bosons and light and heavy-flavor
jets is simulated using ALPGEN; diboson production is simulated with
SHERPA~\cite{Gleisberg:2008ta}.

\begin{figure}[t]
\includegraphics[width=\columnwidth,trim=14 0 8 20,clip=true]{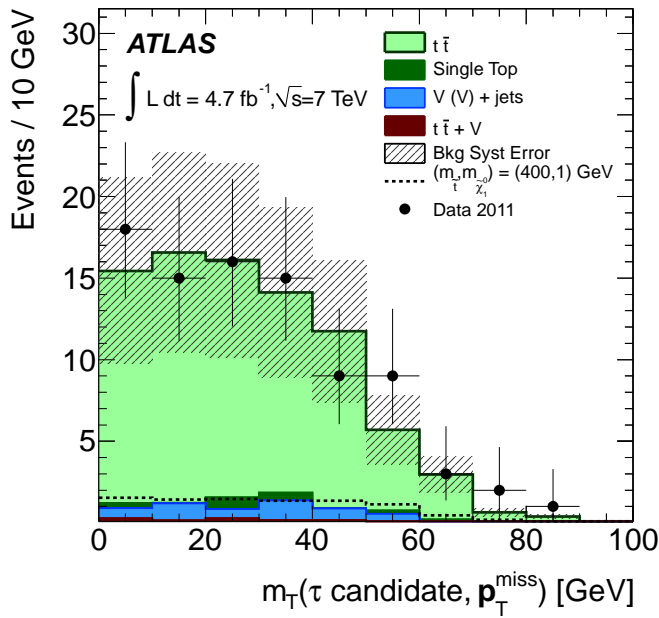}
\caption{\label{fig:tau} 
  The \mt\ distribution for $\tau$-like jets with the selection described in the text.
  Data are indicated by points; shaded histograms
  represent contributions from several SM sources (with \ttbar\ scaled by
  $0.66$).  The hatched error bar indicates the systematic
  uncertainty on the total expected background.  The expected signal
  distribution for $m_{\stopone} = 400\gev, m_{\ninoone} = 1\gev$ is
  overlaid.
}
\end{figure} 
All samples are passed through the Geant4 simulation of the ATLAS
detector, and are reconstructed in the same manner as the data.  The
simulation includes the effect of multiple $pp$ interactions and is
weighted to reproduce the observed distribution of the number of
interactions per bunch crossing.  SM event samples are normalized to
the results of higher-order calculations using the cross sections
cited in Ref.~\cite{Aad:2012ah} except for the $\ell$+jets \ttbar\
background.  This sample is renormalized by a factor that scales the
\ttbar\ expectation to agree with the observed data in a control
region (CR) kinematically close to the signal region but with little
expected signal.  The CR is constructed from events containing one
muon or one ``tight'' electron~\cite{Aad:2011mk} with $\pt>30$~\GeV\
consistent with originating from a $W$-boson decay
($40<\mt^{\ell}<120$~\GeV) and $\geq5$ jets, where $\mt^{\ell}$ is the
transverse mass of the electron or muon and \ptmiss.  The lepton must
be isolated such that the scalar \pt\ sum of tracks in a cone of
$\Delta R<0.2$ around the lepton, excluding the track of the lepton,
is $<1.8$~\GeV\ for the muon or is $<10\%$ of the electron \pt,
respectively.  The jet, $b$-jet, and \MET\ requirements remain the
same as the standard signal selection; however, some topological
constraints are relaxed ($\Delta\phi(\ptmiss,{\rm jet})>\pi/10$
radians and $m_{jjj}<600$~\GeV) and others removed (\mtj) to gain
statistics.  The \ttbar\ purity in the control region is $>80\%$; the
expected signal contamination is $<3\%$.  The lepton is treated as a
jet of the same energy and momentum, mimicking the effect of the
$\tau$ lepton.  Effects of the additional \MET\ from the $\tau$
neutrino are smaller than the statistical uncertainties.  The scale
factor needed to bring the $\geq6$~jet $\ell$+jets ALPGEN \ttbar\
events into agreement with the data after recalculating all quantities
except \MET\ is $0.66\pm0.05$; the uncertainty quoted here is
statistical only. This scale factor is used in
Figs.~\ref{fig:topmass}$-$\ref{fig:met}.  The normalization is
validated with an orthogonal \ttbar-dominated sample created from SRA
by selecting events with $\tau$-like jets; the requirement on \mtj\ is
removed to increase the sample size.  The \mt\ of $\tau$-like jets is
shown in Fig.~\ref{fig:tau}, where the \ttbar\ sample has been
normalized as described above.  Expectations from the simulation agree
with the data within uncertainties.  Contributions to the signal
region from QCD multi-jet and all-hadronic \ttbar\ production are
estimated with a data-driven technique~\cite{Aad:2011ib}. Jets are
smeared in a low-\MET\ data sample using response functions derived
from control regions dominated by multi-jet events.  Only $0.2\pm0.2$
such events remain in SRA after the full event selection.

\begin{figure}[t]
\includegraphics[width=\columnwidth,trim=14 0 8 20,clip=true]{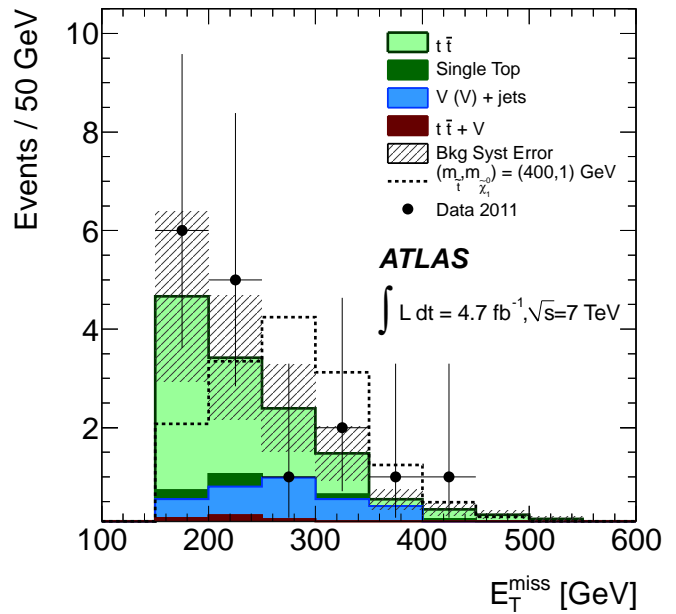}
\caption{\label{fig:met} The \MET\ distribution in data compared to
  the SM expectation for signal region A.  The hatched error bar
  indicates the systematic uncertainty on the total expected
  background.  The expected signal distribution for $m_{\stopone} =
  400\gev, m_{\ninoone} = 1\gev$ is overlaid. The SM background
  distributions do not include the $0.2\pm0.2$ events of all-hadronic
  \ttbar\ and QCD multi-jets estimated from data.}
\end{figure}
The \MET\ distribution in SRA is shown in Fig.~\ref{fig:met} for data,
for the SM backgrounds, and for expectations of \stoppair\ production
with $m_{\stopone}=400$ and $m_{\ninoone}=1$~\GeV.  Numbers of events
and combined statistical and systematic uncertainties, for both SRA
and SRB, are tabulated in Table~\ref{tab:results}.  Uncertainties in
the event generators, including the impact of initial- and final-state
radiation, are the dominant source of systematic uncertainty of 28\%
(23\%) in SRA (SRB).  Other major sources of uncertainty include 22\%
(32\%) for the jet energy calibration, 6.5\% (6.8\%) for jet energy
resolution, 5.9\% (6.2\%) for $b$-jet identification, and 1.4\%
(1.5\%) for \MET\ in SRA (SRB).

The number of observed events in the data is well matched by the SM
background.  These results are interpreted as exclusion limits for
$m_{\stopone}$ and $m_{\ninoone}$ using a CL${}_s$ likelihood ratio
combining Poisson probabilities for signal and
background~\cite{Read:2002hq}. Systematic uncertainties are treated as
nuisance parameters assuming Gaussian distributions.  Uncertainties
associated with jets, $b$-jets, \MET, and luminosity are fully
correlated between signal and background; the others are assumed to be
uncorrelated.  The expected limits for the signal regions are
evaluated for each ($m_{\stopone}$, $m_{\ninoone}$) point; the SR with
the better expected sensitivity is used for that point.  The expected
and observed 95\%~C.L. exclusion limits are displayed in
Fig.~\ref{fig:exclusion}.  Top squark masses between 370~and 465~\GeV\
are excluded for $m_{\ninoone}\sim0$~\GeV\ while
$m_{\stopone}=445$~\GeV\ is excluded for $m_{\ninoone}\le 50$~\GeV.
These values are derived from the $-1\sigma$ observed limit contour to
account for theoretical uncertainties on the SUSY cross sections.  The
95\% CL$_s$ upper limit on the number of events beyond the SM in each
signal region, divided by the integrated luminosity, yields limits on
the observed (expected) visible cross sections of $2.9$ $(2.5)$~fb in
SRA and $1.3$ $(1.3)$~fb in SRB.

\begin{table}[t]
  \caption{
    The numbers of expected events for the SM backgrounds and an example 
    SUSY signal point, and the observed number of events in data. 
    The 95\%~CL${}_s$ upper limit on the observed (expected) 
    visible cross section is appended below.
  }
\label{tab:results}
\begin{center}
\begin{tabular}{lD{.}{.}{-1}@{}l@{~~~~~}D{.}{.}{1.2}@{\hskip 0.03in}l}
\hline\hline
                          & \multicolumn{2}{c}{SRA} & \multicolumn{2}{c}{SRB} \\
\multicolumn{1}{r}{\MET}  & \multicolumn{2}{l}{$>150$~\GeV} & \multicolumn{2}{@{}l}{$>260$~\GeV}  \\ 
\hline\\[-0.9em]
\ttbar                    & 9.2  & $\pm\,2.7$  & 2.3 & $\pm\:\:0.6$ \\[0.1em] 
$\ttbar+\Wboson/\Zboson$  & 0.8  & $\pm\,0.2$  & 0.4 & $\pm\:\:0.1$ \\[0.1em] 
Single top                & 0.7  & $\pm\,0.4$  & 0.2 & \multicolumn{1}{@{}l}{${}^{+\:\:\:\,0.3}_{-\:\:\:\,0.2}$} \\[+0.2em]
$Z+$jets                  & 1.3  & \multicolumn{1}{@{}l}{${}^{+\,\,1.1}_{-\,\,1.0}$}  & 0.9 & \multicolumn{1}{@{}l}{${}^{+\:\:\:\,0.8}_{-\:\:\:\,0.7}$} \\[+0.2em]
$W+$jets                  & 1.2  & \multicolumn{1}{@{}l}{${}^{+\,\,1.4}_{-\,\,1.0}$}  & 0.5 & $\pm\:\:0.4$ \\[+0.2em]
Diboson                   & 0.1  & \multicolumn{1}{@{}l}{${}^{+\,\,0.2}_{-\,\,0.1}$}  & 0.1 & \multicolumn{1}{@{}l}{${}^{+\:\:\:\,0.2}_{-\:\:\:\,0.1}$} \\[+0.2em] 
Multi-jets                & 0.2  & $\pm\,0.2$  & 0.02 & $\pm\:\:0.02$ \\[0.1em] 
\hline\\[-0.9em]
Total SM                  & 13.5 & \multicolumn{1}{@{}l}{${}^{+\,\,3.7}_{-\,\,3.6}$}  & 4.4 & \multicolumn{1}{@{}l}{${}^{+\:\:\:\,1.7}_{-\:\:\:\,1.3}$}  \\[+0.2em] 
SUSY $(m_{\stopone},m_{\ninoone})=(400,1)\GeV$ & 14.8 & $\pm\,4.0$ & 8.9 & $\pm\:\:3.1$ \\
Data (observed)           & \multicolumn{2}{c}{16} & \multicolumn{2}{l@{}}{~~~~~4} \\ 
\hline\\[-1ex]
Visible cross section [fb]& \multicolumn{2}{c}{2.9 (2.5)} & \multicolumn{2}{l@{}}{1.3 (1.3)} \\
\hline\hline
\end{tabular}
\end{center}
\end{table}%
In conclusion, we have presented a search for the direct production of
\stoppair\ in the all-hadronic $t\ninoone \bar{t}\ninoone$ decay
channel, assuming ${\cal B}(\stopone \rightarrow t\ninoone) = 100\%$.
A total of $16\;(4)$ events are observed compared to a predicted
Standard Model background of ${13.5}^{+\,\,3.7}_{-\,\,3.6}\;
({4.4}^{+\,\,1.7}_{-\,\,1.3})$ events in two signal regions based on
\ensuremath{\int\!\!{\cal L}\,{\rm d}t = 4.7\,\ifb} of $pp$ collision
data taken at $\rts=7\tev$ with the ATLAS detector at the LHC.  No
evidence for \stoppair\ is observed in data and 95\%~C.L. limits are
set on $\stoppair$ production as a function of $m_{\stopone}$ and
$m_{\ninoone}$.
\begin{figure}[t]
\includegraphics[width=\columnwidth,trim=0 0 10 0,clip=true]{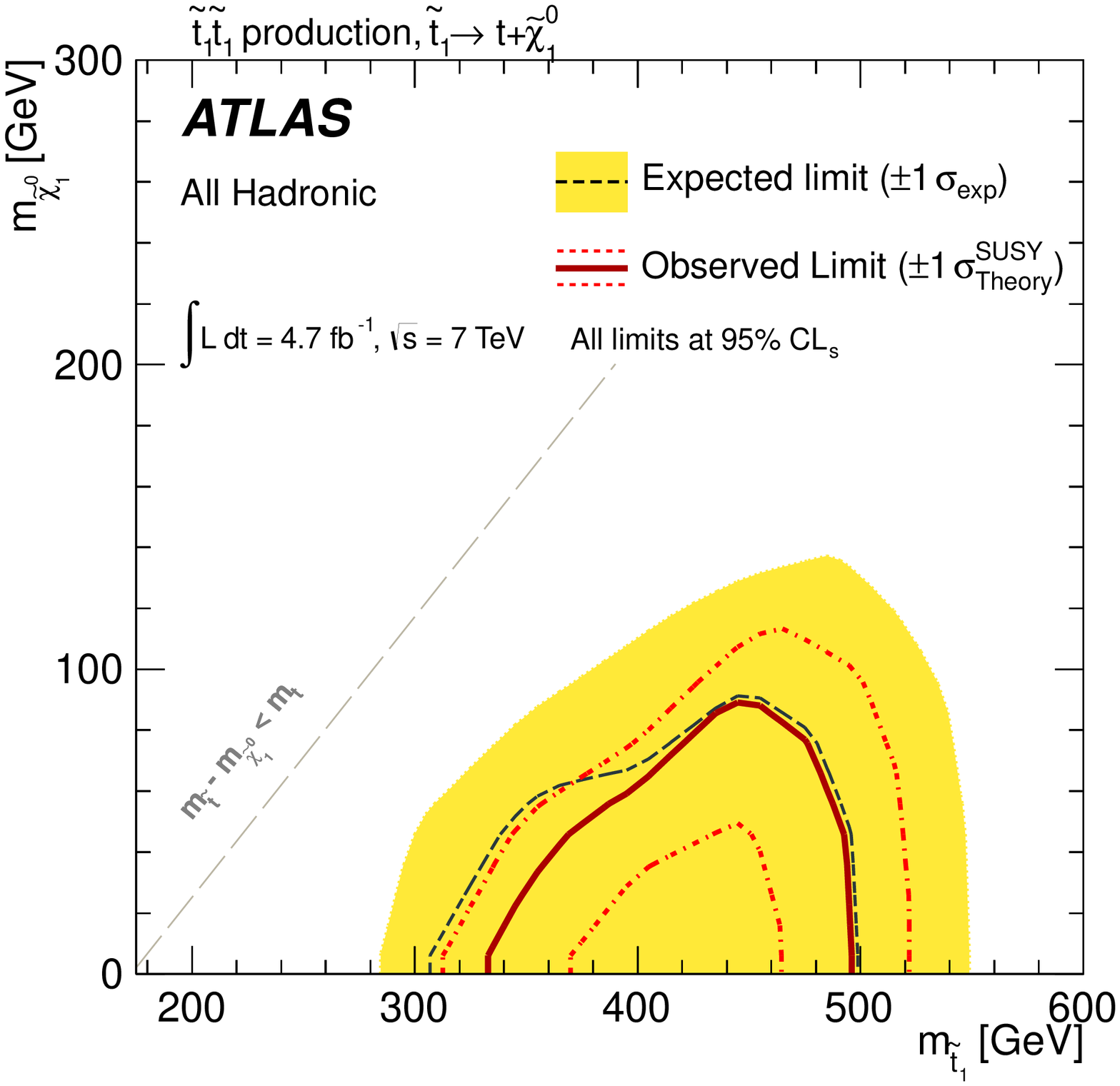}
\caption{\label{fig:exclusion} Expected and observed 95\% CL${}_s$
  exclusion limits in the plane of $m_{\ninoone}$ vs. $m_{\stopone}$,
  assuming ${\cal B}(\stopone \rightarrow t\ninoone) = 100\%$.  The
  dashed line shows the expected limit at 95\% C.L.  with the shaded
  region indicating the $\pm1\sigma$ exclusions due to experimental
  uncertainties.  Observed limits are indicated by the solid contour
  (nominal) and the dotted contours (obtained by varying the SUSY
  cross section by the theoretical uncertainties).  The inner dotted
  contour indicates the excluded
  region.} 
\end{figure}

\begin{acknowledgments}
We thank CERN for the very successful operation of the LHC, as well as the
support staff from our institutions without whom ATLAS could not be
operated efficiently.

We acknowledge the support of ANPCyT, Argentina; YerPhI, Armenia; ARC,
Australia; BMWF, Austria; ANAS, Azerbaijan; SSTC, Belarus; CNPq and FAPESP,
Brazil; NSERC, NRC and CFI, Canada; CERN; CONICYT, Chile; CAS, MOST and NSFC,
China; COLCIENCIAS, Colombia; MSMT CR, MPO CR and VSC CR, Czech Republic;
DNRF, DNSRC and Lundbeck Foundation, Denmark; EPLANET and ERC, European Union;
IN2P3-CNRS, CEA-DSM/IRFU, France; GNAS, Georgia; BMBF, DFG, HGF, MPG and AvH
Foundation, Germany; GSRT, Greece; ISF, MINERVA, GIF, DIP and Benoziyo Center,
Israel; INFN, Italy; MEXT and JSPS, Japan; CNRST, Morocco; FOM and NWO,
Netherlands; RCN, Norway; MNiSW, Poland; GRICES and FCT, Portugal; MERYS
(MECTS), Romania; MES of Russia and ROSATOM, Russian Federation; JINR; MSTD,
Serbia; MSSR, Slovakia; ARRS and MVZT, Slovenia; DST/NRF, South Africa;
MICINN, Spain; SRC and Wallenberg Foundation, Sweden; SER, SNSF and Cantons of
Bern and Geneva, Switzerland; NSC, Taiwan; TAEK, Turkey; STFC, the Royal
Society and Leverhulme Trust, United Kingdom; DOE and NSF, United States of
America.

The crucial computing support from all WLCG partners is acknowledged
gratefully, in particular from CERN and the ATLAS Tier-1 facilities at
TRIUMF (Canada), NDGF (Denmark, Norway, Sweden), CC-IN2P3 (France),
KIT/GridKA (Germany), INFN-CNAF (Italy), NL-T1 (Netherlands), PIC (Spain),
ASGC (Taiwan), RAL (UK) and BNL (USA) and in the Tier-2 facilities
worldwide.
\end{acknowledgments}

\bibliography{paper}
\onecolumngrid

\clearpage
% ATLAS Collaboration author list
% Data extracted on 12-Jul-2012 for paper reference SUSY-2012-05
%\documentclass[11pt]{article}
%\usepackage{a4wide}\begin{document}
\begin{flushleft}
{\Large The ATLAS Collaboration}

\bigskip

G.~Aad$^{\rm 48}$,
T.~Abajyan$^{\rm 21}$,
B.~Abbott$^{\rm 111}$,
J.~Abdallah$^{\rm 12}$,
S.~Abdel~Khalek$^{\rm 115}$,
A.A.~Abdelalim$^{\rm 49}$,
O.~Abdinov$^{\rm 11}$,
R.~Aben$^{\rm 105}$,
B.~Abi$^{\rm 112}$,
M.~Abolins$^{\rm 88}$
% ...
\bigskip

% ...
$^{11}$ Institute of Physics, Azerbaijan Academy of Sciences, Baku, Azerbaijan\\
$^{12}$ Institut de F\'{i}sica d'Altes Energies and Departament de F\'{i}sica de la Universitat Aut\`{o}noma de Barcelona and ICREA, Barcelona, Spain\\
% ...
$^{21}$ Physikalisches Institut, University of Bonn, Bonn, Germany\\
% ...
$^{48}$ Fakult\"{a}t f\"{u}r Mathematik und Physik, Albert-Ludwigs-Universit\"{a}t, Freiburg, Germany\\
$^{49}$ Section de Physique, Universit\'{e} de Gen\`{e}ve, Geneva, Switzerland\\
% ...
$^{88}$ Department of Physics and Astronomy, Michigan State University, East Lansing MI, United States of America\\
% ...
$^{105}$ Nikhef National Institute for Subatomic Physics and University of Amsterdam, Amsterdam, Netherlands\\
% ...
$^{111}$ Homer L. Dodge Department of Physics and Astronomy, University of Oklahoma, Norman OK, United States of America\\
$^{112}$ Department of Physics, Oklahoma State University, Stillwater OK, United States of America\\
% ...
$^{115}$ LAL, Universit\'{e} Paris-Sud and CNRS/IN2P3, Orsay, France\\
% ...
\end{flushleft}

%\end{document}
% Created with TexConverter

\end{document}